# INVESTIGATION OF THE CELLULAR RESPONSE TO BONE FRACTURES: EVIDENCE FOR FLEXOELECTRICITY

Raquel Núñez-Toldrà [a,b,¥,*], Fabian Vasquez-Sancho [a,c,¥], Nathalie Barroca [a], Gustau Catalan [a,d,*]

[a] Institut Catala de Nanociencia i Nanotecnologia (ICN2), CSIC and the Barcelona Institute of Nanoscience and Nanotecnology (BIST), Bellaterra 08193, Barcelona, Catalonia.

[b] National Heart & Lung Institute, Imperial College London, London W12 0NN, UK

[c] Centro de Investigación en Ciencia e Ingeniería de Materiales, Universidad de Costa Rica, San José 11501, Costa Rica

[d] Institució Catalana de Recerca i Estudis Avançats (ICREA), Barcelona, Catalonia.

[¥] These authors contributed equally

**\*Corresponding authors at:**
Institut Catala de Nanociencia i Nanotecnologia (ICN2), CSIC and the Barcelona Institute of Nanoscience and Nanotecnology (BIST). Bellaterra 08193, Barcelona
E-mail: r.nunez-toldra@imperial.ac.uk (Raquel Nuñez-Toldra); gustau.catalan@icn2.cat (Gustau Catalan)

## Abstract

The recent discovery of bone flexoelectricity (strain-gradient-induced electrical polarization) suggests that flexoelectricity could have physiological effects in bones, and specifically near bone fractures, where flexoelectricity is theoretically highest. Here, we report a cytological study of the interaction between crack stress and bone cells. We have cultured MC3T3-E1 mouse osteoblastic cells in biomimetic microcracked hydroxyapatite substrates, differentiated into osteocytes and applied a strain gradient to the samples. The results show a strong apoptotic cellular response, whereby mechanical stimulation causes those cells near the crack to die, as indicated by live-dead and caspase staining. In addition, analysis two weeks post-stimulation shows increased cell attachment and mineralization around microcracks and a higher expression of osteocalcin –an osteogenic protein known to be promoted by physical exercise.

The results are consistent with flexoelectricity playing at least two different roles in bone remodelling: apoptotic trigger of the repair protocol, and electro-stimulant of the bone-building activity of osteoblasts.

**Introduction**

Microcracks are a ubiquitous feature of all bones, the result of fatigue damage due to recurrent and cyclically applied stresses [1-3]. They are common in horses, greyhounds and human athletes [4-6], and they act as seeds of bone remodelling but also, under very intense loadings, as sources of mechanical fatigue [5,7]. Mechanically, microcracks may also play a role in preventing bone failure by diverting and dissipating strain energy of larger cracks, thus improving the overall bone toughness [8]. The physiological process of local bone remodelling in response to cracking involves the participation of three main cell types: osteoblasts, osteocytes and osteoclasts [2]. Osteocytes, which are embedded in the bone matrix, establish an extensive intracellular and extracellular communication system via gap-junction-coupled cell processes and canaliculi. The process starts with the injury of some osteocytes, which, upon dying, release chemical signals that promote the recruitment of osteoclasts for the resorption phase (cleaning of the damaged area) [9,10]. Osteoblasts then initiate new bone formation with the creation of the unmineralized extracellular matrix that successively becomes calcified. Furthermore, the generation of a microcrack also promotes the maturation of the osteoblasts to osteocytes through the release of different growth factors (e.g. TGF-b, BMPs, IGFs, and PDGF) [11].

The cytological activity of cracks can, at least in part, be traced to mechano-chemical effects. Microcracking increases mineral delivery (principally calcium and inorganic phosphate ions) into the medium [11,12], having a strong effect on bone cells, stimulating osteoblast growth, differentiation and matrix mineralization *in vitro* [11-14]. In addition, calcium ion efflux from bone matrix appears in notched bone samples subject to stress, being maximally concentrated near the apex of the notches, where stress and stress gradients are largest [11]. The actual link between mechanical stress and bio-chemical effects, however, remains to be clarified. This is where the electromechanical properties of bones (their ability to generate electric fields in response to mechanical stress) may play a role.

Bone remodelling has been related to electromechanics since Fukada and Yasuda's seminal discovery of piezoelectricity in bone tissue, attributed to the presence of piezoelectric collagen [15-17]. Recently, however, it has been discovered that bones' ability to generate voltage in response to mechanical stress is retained even in the absence of collagen –as long as the applied deformation is inhomogeneous– thanks to flexoeletricity [18,19]. Flexoelectricity is a general property that allows materials of any symmetry (including non-piezoelectric ones) to generate a voltage in response to strain gradients [19,20].

The discovery of flexoelectricity in bone and bone mineral (hydroaxyapatite, HA) [18] has the potential to affect our present understanding of bone physiology. The structure of bones –and thus their distribution of mechanical stresses- is inherently inhomogeneous, so flexoelectricity is an inevitable response to any mechanical loading. Moreover, bone damage in the form of micro-fractures also creates mechanical inhomogeneities with large local strain gradients that can generate strong flexoelectric signals. Theoretical calculations indicate that flexoelectric fields around bone microcracks can reach several kV/m near the crack apex [18], and this is of the same order of magnitude as the electrostatic fields that are known to induce osteocyte apoptosis: as shown in *in vitro* experiments, 1 kV/m already can damage cells and fields higher than 10 kV/m cause their immediate necrosis [21]. Flexoelectricity can in theory generate fields of such magnitude within tens of microns of a crack [18], and there is experimental evidence for even more intense fields within smaller (nanoscopic) distances [22]. Since crack repair begins with osteocyte apoptosis [9], an apoptotic role of crack flexoelectricity would have profound implications for bone remodelling.

It is the purpose of this work to examine experimentally the cellular response to mechanical stress in cracked bone mineral and determine whether cytological effects can be linked to flexoelectricity. To this end, we cultured osteoblastic cells onto biomimetic cracked HA and TiO$_2$ substrates; the use of these different materials, both of which are flexoelectric and not piezoelectric, ensured the generality of the results and facilitated the flexoelectric (as opposed to piezoelectric) analysis. Strain gradients were generated by applying a bending stress to the substrates, which was locally concentrated by the cracks. The flexoelectric field generated by these strain gradients was calculated using standard elastic theory and experimentally measured values of the flexoeoectric coefficient (details are provided in the supplementary materials). Its effect at the cellular level was analysed both immediately and after 12 days of osteogenic differentiation post-bending (Scheme 1).

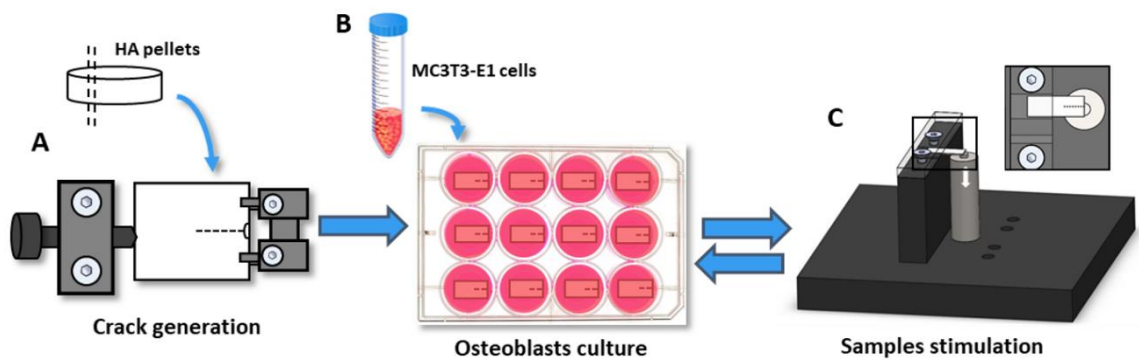

**Scheme 1. Schematic illustration of the experiment design.** A) Sample preparation and crack generation by a three-point bending system. A notch was created, then a load was applied parallel to the width of the sample with a sharp knife drived by a micrometre screw. B) MC3T3-E1 mouse osteoblasts culture on HA and TiO$_2$ pre-cracked samples. C) Samples stimulation mechanism. A compressive strain on the top surface was induced using a preloaded piezoelectric actuator to stimulate the samples. 50 N were applied at 5 Hz for 150s.

## Results

**Osteocyte apoptosis around bone microcracks**

In order to evaluate the survival of the osteocytes after crack generation, cell viability was analysed by Live/Dead staining before and after cracking in HA substrates with MC3T3-E1 osteoblasts cultured on the surface (Figure 1). Figure 1A shows that, before cracking, cells were well attached to the surface and no dead cells were detected. However, immediately after crack generation (Figure 1B), high levels of cell damage around the crack were observed. Dead cells were detected up to a distance of 200 µm from the crack, while the rest of the osteocytes survived the event, demonstrating a direct causal link between cracking and cell death. Although there are dead cells along the entire crack, we notice that the effect is higher around the notch. This is consistent with the high concentration of stress at the notch, which acts as the seed of the crack. This –high stress is partially relieved by the appearance of the crack, which is relevant because flexoelectric fields are proportional to differences in strai

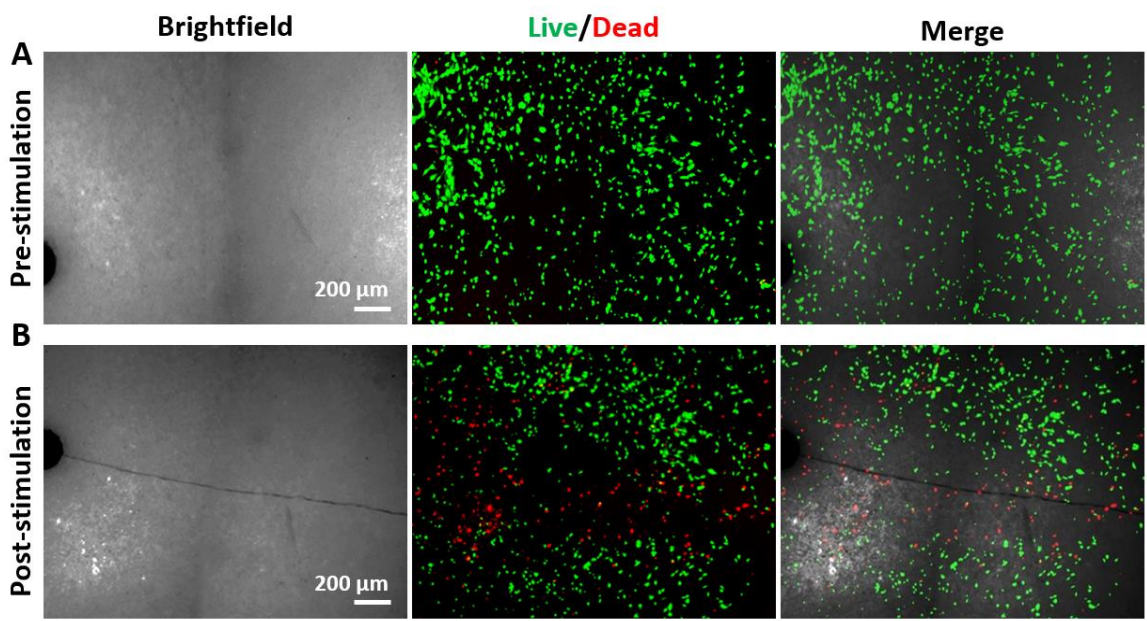

**Figure 1. Osteocytes' reaction to cracking.** Cell viability analysed by Live/Dead staining (A) before and (B) immediately after crack generation in HA substrate.

**Apoptotic effects of crack-generated flexoelectricity on osteocytes**

Having observed that the appearance of a crack has an apoptotic effect, the next question is how this effect is achieved. The samples are non-porous, polished ceramics, so streaming potentials and fluid flow disruption by the crack, which can be an issue in real bones, can in principle be discarded from the analysis of the present results. This still leaves other explanations: mechanoelectric (flexoelectric field emanating from the crack), mechanochemical (debris released by the crack) or plain mechanical (shock-wave generated by the cracking event). To discard these mechanisms, MC3T3-E1 cells were cultured in pre-cracked and cleaned HA substrates, so that the potential effects of the cracking event (mechanical shock and/or release of debris) could be removed.

After 10 days in osteogenic medium, the pre-cracked samples were mechanically stimulated (bent) and analysed before and immediately after mechanical stimulation (Figure 2). The stimulation consisted in periodically bending the samples with an oscillatory force for 150 seconds at a frequency of 5 Hz using the rig depicted in Scheme 1, so that the cracks undergo an oscillatory elastic deformation, thereby generating an alternating flexoelectric field.

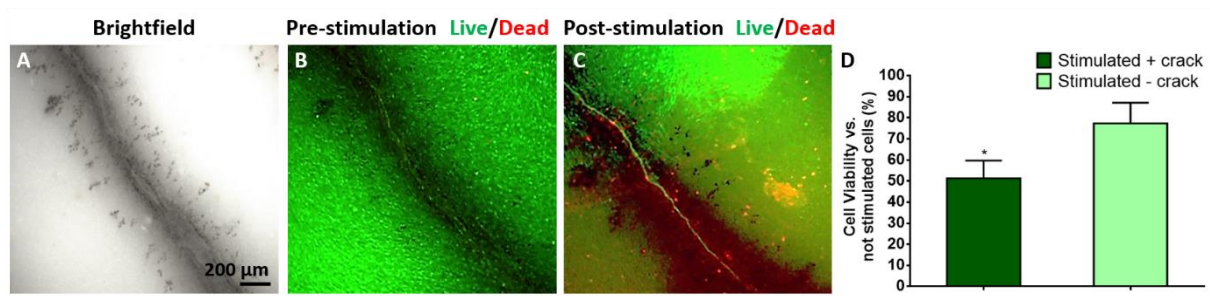

**Figure 2. Osteocytes' reaction to mechanical stimulation of already existing cracks.** A) Brightfield optical micrograph of the crack prior to culture. B) Live/Dead staining image of the culture on the crack before bending the substrate. C) Live/Dead staining image after bending the substrate. D) Cell viability quantification after osteocytes stimulation by MTT assay versus not stimulated samples. Mechanical stimulation barely affects cell viability when there is no crack ('stimulated - crack') whereas it has a strong effect when there is a crack ('stimulated + crack'). (N=6). *$P<0.05$.

Live/Dead staining (Figure 2B) shows that, in pre-cracked samples, the cells are initially alive across the entire sample, including the vicinity of the crack. This shows that the mere presence of the crack does not cause apoptosis: the crack must be mechanically stimulated for cells to die. Immediately after mechanically stimulating the crack by bending the sample, however, a dramatic "death trail" appears along the crack (Figure 2C). Apoptosis progression in osteocytes after bending was additionally detected by Caspase 3/7 staining in green (Figure 3). The maximum levels of apoptosis were detected from the middle to the tip of the crack, consistent with the higher strain gradient near the tip.

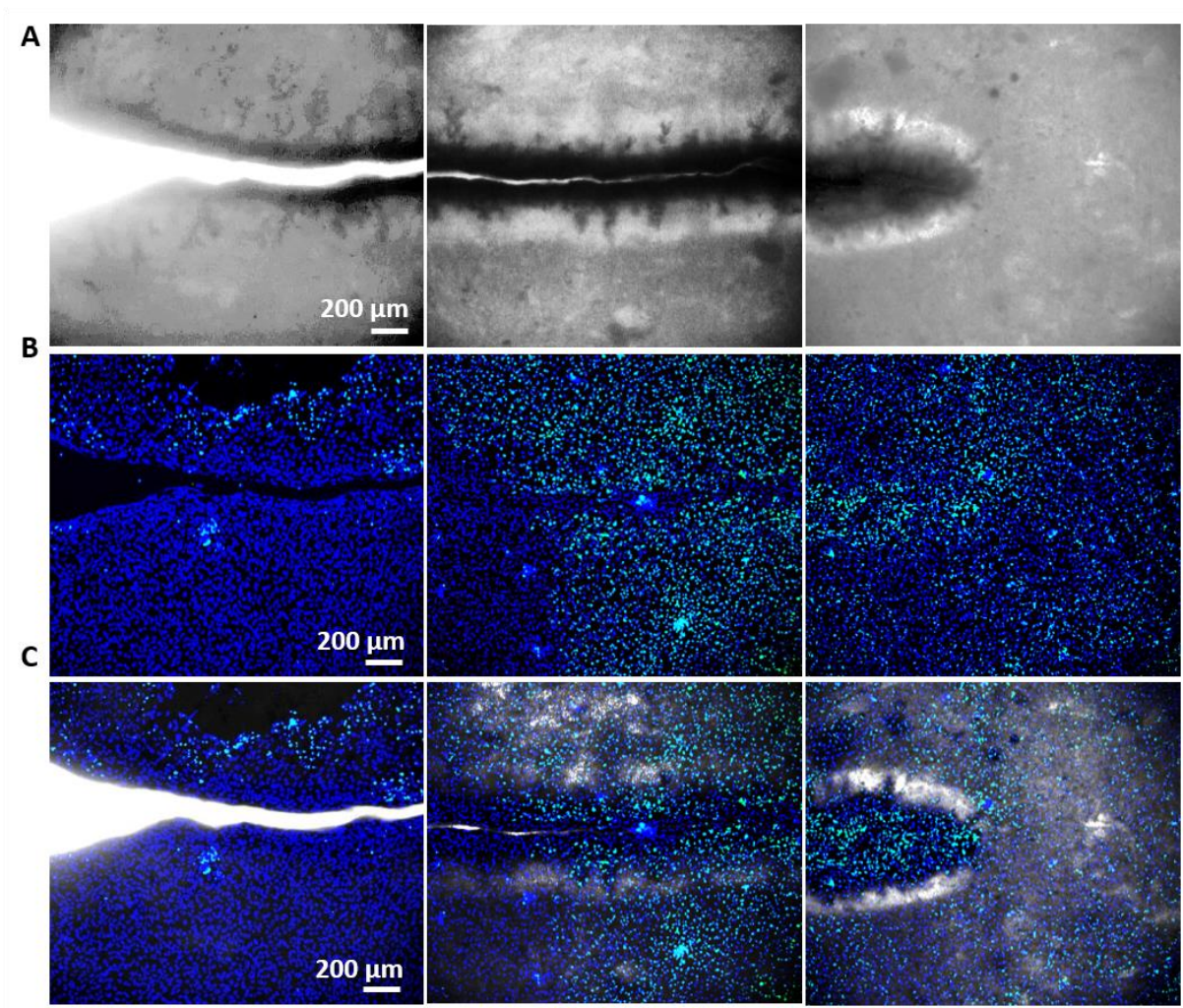

**Figure 3. Caspase-3/7 activity under flexoelectric stimulation of hydroxyapatite ceramics.** A) Brightfield scannint electron image of the base, centre and tip of the crack. B) Live/Dead staining images. C) merged images of the crack and the osteocytes post. Caspase-3/7 activity

was assayed by the CellEvent caspase-3/7 detection reagent to detect apoptosis in turquoise green and Hoechst 33342 nucleic acid stain to label nuclei in dark blue.

The cell viability in osteocytes after crack bending was quantified by a cell proliferation assay and compared to a reference sample that was bent but had no crack. The results show that around 75% of the osteocytes were viable after bending the uncracked reference, whereas cell viability after bending cracked samples was significantly decreased to around 50% (Figure 2D). Notice that these results are an underestimate of the local effect of the crack, since they average over the entire sample surface; the microscopy images show that, locally, the results are more radical, with virtually 0% survival rate for cells within 100 µm of the crack. These results therefore rule out mechanical shock-wave effects and any other effect due to the appearance of the crack, since the cracking had been *before* cell culture. The results also rule out the effect of debris release, since the pre-cracked samples had been cleaned before cell culture.

These experiments were complemented using a different substrate material, titanium dioxide ($TiO_2$), with different elastic and flexoelectric coefficients. Like HA, this material is flexoelectric and not piezoelectric, but it differs from HA in its theoretical spatial distribution of flexoelectricity. Specifically, according to the numerical calculations (see Appendix), cracks in $TiO_2$ should have a smaller radius of action (Figure 4B1). Experimentally, the apoptotic effect of crack flexoelectricity was observed to be indeed limited to cells immediately adjacent to the crack (Figure 4B2), whereas for HA the apoptotic region was much larger (Figure 4A2).

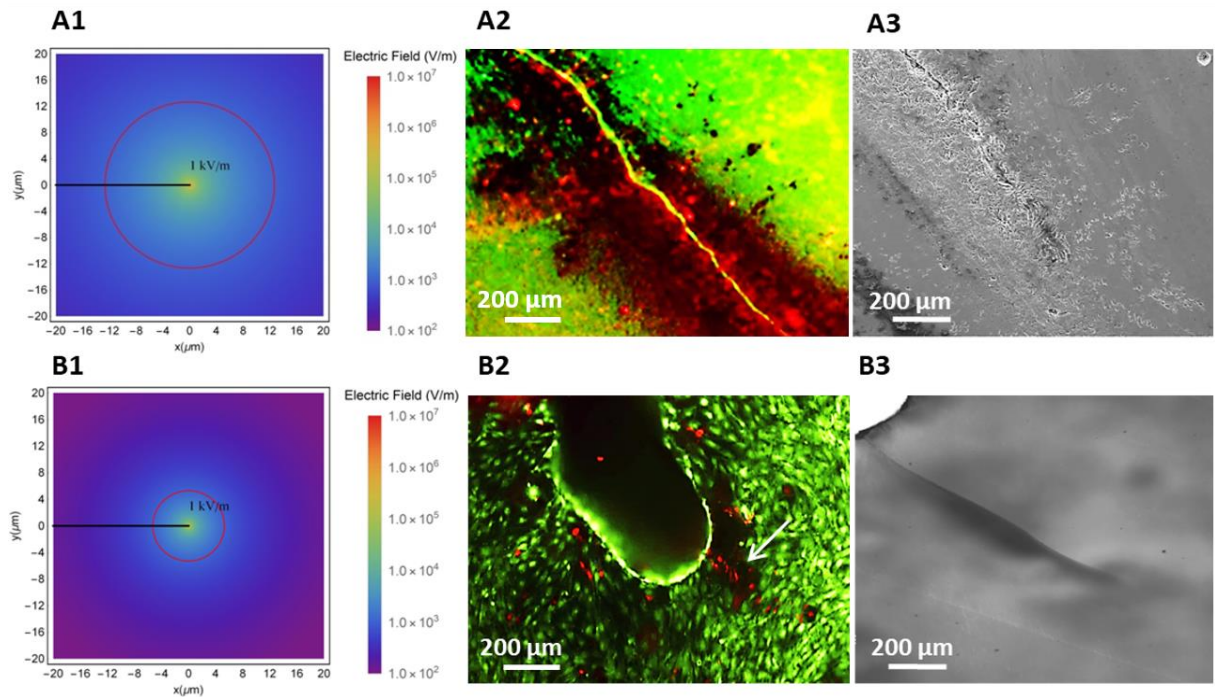

**Figure 4. Flexoelectricity effects around bone microcracks in different materials.** Calculated flexoelectric field distribution around a microcrack in (A1) HA and (B1) TiO$_2$. Red circle limits the 1 kV/m region, the minimum magnitude that can induce damage to cells [21]. Live/Dead staining images of osteocytes cultured in (A2) HA and (B2) TiO$_2$ immediately after stimulation. Arrows indicate the localization of the cracks in all the materials. Images of microcracks prior cell culture in (A3) HA and (B3) TiO$_2$ substrates.

On the other hand, it has to be noted that the morphology of the TiO$_2$ crystals was different from that of the HA ceramics, and this is relevant for rationalizing the extension of the "damage zone". In HA, cell apoptosis is observed at distances larger than theoretically calculated (Figure 4A1), and we believe that this is due to the presence of subsidiary microcracks and mechanical damage around the main fissure, as observed with high-resolution SEM image (Figure 4A3); in contrast, the crack in TiO$_2$ single crystal was clean and had no branches (Figure 4B3), so it is a truer embodiment of the ideal case simulated by the calculations. Both the different material and the different crack morphology contribute to the final apoptotic range, and without HA single crystals we cannot deconvolute these two factors.

The take-home messages from the analysis so far are that (i) all experiments were using compact, collagen-free and non-piezoelectric materials, so neither piezoelectricity nor streaming currents can be invoked as a source of mechano-electrical activity, and (ii) cell viability is not affected by the mere presence of the crack (as shown by Figure 2B), nor by bending uncracked samples (Figure 2D): we need both the presence of the crack AND its mechanical stimulation in order to achieve the apoptotic effect.

**Mineralizing effects of fracture stimulation**

The results in the previous section are consistent with flexoelectricity as a cause for osteocyte apoptosis –the initial step of the repair process [9]. However, the healing of a crack ultimately requires the bone-forming action of osteoblasts. It is thus important to examine whether the mineralizing activity of these cells may also be affected by cracks –and, if so, whether the link is flexoelectric.

To separate out the effect of ionic efflux (i.e. calcium ions being ejected when the crack is formed) from that of flexoelectricity, we chose again to culture osteoblastic cells on pre-cracked HA samples. The samples were then mechanically stimulated at day 3 of culture, using the same protocol as in the apoptosis experiment (150 seconds at 5 Hz). Then, the samples were analysed after 15 days of osteogenic culture (12 days post-bending). Mineralization and cell survival were evaluated simultaneously in stimulated and control (cracked but not bent) samples. The results show that the cell viability was similar for both sets, with noticeably higher cell population near the crack in both cases. Osteoblast growth is thus favoured by the presence of the crack even in the absence of mechanical stimulation, indicating that texture by itself is a positive factor irrespective of flexoelectric activity.

Mineralization is also favoured by the presence of cracks, with Xylenol Orange staining showing that significantly more mineral deposition occurred near the cracks, including those that had not been mechanically stimulated (Figure 5A). This result again suggests that the topographic disruption (i.e. texture change) associated with the crack is already capable of stimulating the mineralizing activity of the cells. On the other hand, the mechanically stimulated samples do show more mineral deposition and higher expression levels of osteocalcin (Figure 5B), which is consistent with an additional positive role of flexoelectricity.

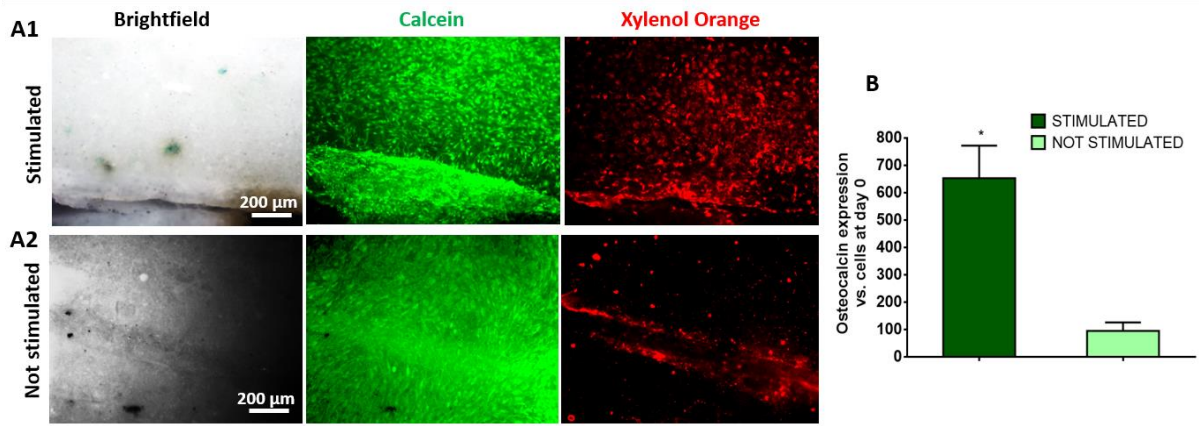

**Figure 5. Osteoblasts mineralization by flexoelectricity.** A) Brightfield, Calcein and Xylenol Orange staining images of osteocytes cultured in HA samples over 15 days in (A1) stimulated and (A2) not stimulated samples. B) OC expression after 15 days of osteogenic induction in stimulated and not stimulated HA samples (N=6).*P<0.05*.

**Discussion**

The results show that mechanical loading of micro-fractures has two effects on bone cells: apoptosis of the osteocytes near the crack, and increased maturation and mineralization of osteoblasts over the subsequent differentiation days (Scheme 2). Cell apoptosis was observed along the length of the crack up to a distance of 200 μm away from it. These results are in line with results for natural bone, yet we have observed them in compact, collagen-free ceramics and crystals. This indicates that the apoptotic effect of cracks does not require the presence of piezoelectric collagen nor streaming currents [9,23], at least *in vitro*, and flexoelectricity is in theory sufficient to account for the results.

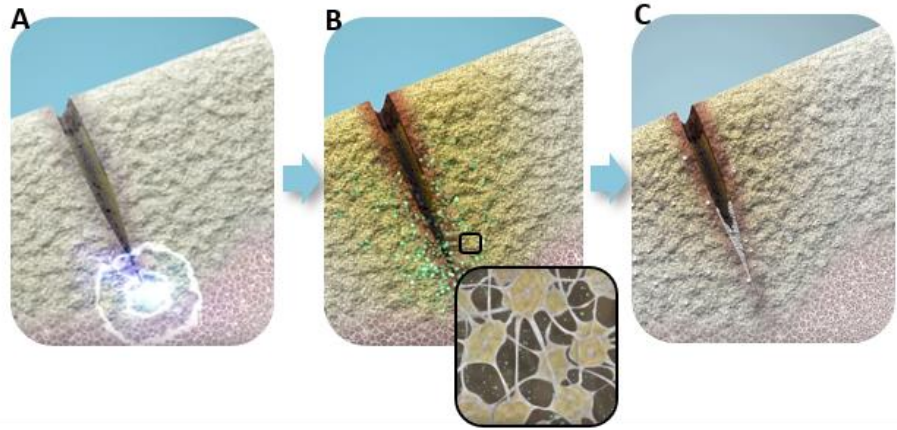

**Scheme 2. Schematic of the role of flexoelectricity in bone repair and remodelling.** Crack-generated flexoelectricity (A) is theoretically large enough to induce osteocyte apoptosis (B), and increase osteoblasts migration and mineralization around these microcracks (C) initiating the crack- healing process.

These result *per se*, however, would not rule out other mechanical effects. The osteocyte network is a mechanosensory system suitable for mechanotransduction. When a crack is generated, there is an increased signalling for apoptosis by chemotactic and mechanotransduction signals within the osteocyte network [10,24]. On the other hand, the small magnitude of the strains induced by mechanical stimulation in our bending experiment on already-cracked samples (<1%) are too small to cause by itself the immediate death of cells, which are flexible entities capable of withstanding large deformations. Also, as discussed, the experiments on pre-cracked and cleaned substrates rule out other effects as the shock-wave generated by the fracture event, the release of crack-generated debris, or indeed the physical ripping apart of cells spanning the two sides of the crack as it appears, since the cracks are already in place before the cells are grown instead of the other way round.

The apoptotic effect of mechanical stimulation of cracks was also evaluated on single crystal substrates of titania ($TiO_2$) which, like hydroxyapatite, is flexoelectric and not piezoelectric. The experimental results on $TiO_2$ were consistent with flexoelectric expectations for flexoelectricity, namely (i) mechanically stimulated cracks induce apoptosis and (ii) the apoptosis was confined within a much smaller distance of the crack than hydroxyapatite. However, the latter result is not conclusive because, besides the smaller theoretical radius of flexoelectricity, there can be a second factor at play which is the morphology of the crack – a clean cleavage for $TiO_2$ crystals compared to the more irregular and branched crack in HA ceramics.

Unfortunately, it is not technically possible for us to directly measure the flexoelectric fields as this would require placing tiny electrodes at microscopic distances of the crack apex on an oscillating substrate covered in living cells. The challenge of directly measuring fracture flexoelectricity is formidable even *in vitro* and forbidding *in vivo*. Nevertheless, the present results provide compelling evidence for a physiological role of flexoelectricity as a cause of osteocyte apoptosis, and thus a trigger mechanism for bone repair in microdamaged bone mineral [9,25]. Moreover, as the crack is healed and the tip recedes, the focus of highest flexoelectricity must move with it. Flexoelectricity would thus act not only as the starting shot but also as a Parthian shot trailing the crack apex.

Of course, if cracks were constantly 'exercised', the same electric fields that cause apoptosis would also impede the regenerative activity of osteoblasts and osteoclasts –put bluntly, these cells would also get electrocuted. We therefore hypothesise that the alternation between activity and rest is important for a successful remodelling.

With this in mind, we have also examined a potential impact of sporadic flexoelectricity on mineralization. The maturation and mineralization of osteoblasts was studied by stimulating osteoblast cultures on pre-cracked samples and analysing the results after 12 days of osteogenic culture without further stimulation. The results show that cracks stimulate mineralization even when there is no mechanical stimulus, indicating an important role of texture by itself, but there is some evidence that the mechanical loading of the cracks promotes additional osteoblastic activity. In particular, stimulated microcracks display increased cell attachment and mineralization and higher expression of osteocalcin. The latter is a genetic marker related to osteogenesis whose presence in blood increases with physical exercise [26]. These results hence provide a potential physical explanation (or at least a contributing factor) for the observed link between physical exercise and increased generation of osteocalcin.

Different studies have shown that electrical stimuli can affect the properties and the regenerative capacity of skeletal tissues [27,28]. Osteoblasts in culture are sensitive to electrical stimulation, in which causes an enhancement of their differentiation and mineralization capacities [21]. Electric fields also originate disruption or alteration of ionic gradients or cell surface charges, leading to changes in cell signalling pathways and gene expression [29,30]. Cells respond to electric stimulation by the activation of calcium channels of the cell membrane and release of calcium from intracellular calcium stores [31]. Low-frequency electric fields significantly affect stem cell populations, by enhancing cell signalling pathways and differentiation [21,32]. The stimulating effect of external electric fields on bone remodelling is thus well documented. Although other mechanisms may also contribute *in vivo*, the present work suggests that the bones themselves, via crack-generated flexoelectricity, can provide their own repairing electrical stimulation.

**Methods**

*Substrate materials*

Bovine femurs were obtained directly from the slaughter house. We produced our own HA samples from bovine bones following the procedure of Ooi, C. et al. [33] After annealing the bones at 900 ºC during 2 h of holding time (ramps of 5 ºC/min), they were free from organic matter had a bleached-white aspect. In order to make the compact discs, they were milled and the powder was sieved to 125 µm particle size. The powder was then uniaxially pressed into pellets of 22.5 mm of diameter under a pressure 25 metric tons. Finally, the pellets were air sintered at 1360º for 4 hours. Pellets were cut using a diamond wire at low speed in order to avoid damage. Both surfaces of the samples were polished up to 1 µm grain size discs with an Allied precision polishing system at low velocity to minimize damage to the samples. Supplementary Figure 1 shows the X-ray diffraction confirming the hydroxyapatite structure of the resulting ceramic pellets. The $TiO_2$ samples were one-side-polished monocrystalline substrates (rutile structure) commercially acquired from CrysTec.

*Cleaning protocol of substrates*

Before cell culture, all samples were put in the oven for 12 h at 900 °C. Then, the samples were immerged in ethanol for 15 minutes. Finally, the substrates were dry-air while being exposed to UV light for 15 minutes.

*Crack generation*

Samples with cells (Figure 1) and without cells (Figures 2-5) were cracked by using a three-point bending system with a sharp stainless steel knife driven by a micrometer screw as can be seen in Scheme 1A. Cracking a sample with cells (Figure 1) shows the effect of the appearance of cracks. Growing cells on an already-cracked sample (Figures 2-5) shows the effect of the existence of cracks. A notch was created to the samples, then the load was applied parallel to the width of the sample and the system was designed for samples of different thicknesses, lengths and widths. The common dimensions for these samples were 12x6x1 mm.

*Cell Culture*

MC3T3-E1 mouse osteoblastic cells (ECACC General Cell Collection) were seeded at 0.5 -2 x10,000 cells/cm² and expanded in MEMα GlutaMAX medium (Gibco), supplemented with

10% foetal bovine serum (FBS, Life Technologies) and 1% penicillin/streptomycin (Life Technologies) at 37°C in a 5% $CO_2$ incubator. Then, for osteogenic differentiation, cells were cultured on HA and $TiO_2$ surfaces at 11.300 cells/$cm^2$ and the medium was further supplemented with 8 mM β-glycerol phosphate (Sigma-Aldrich), 50 μg/ml L-ascorbic acid (Sigma-Aldrich) and 10 nM Dexamethasone (Sigma-Aldrich).

*Sample Stimulation*

Cell cultures were grown on cantilever-shaped ceramic samples with a crack running half way along the longitudinal direction. The cantilevers were rigidly clamped at one end while the free end of the cantilever was periodically pushed with a piezoelectric actuator (PiMicos) model P-820.10, delivering an oscillating upward force of 50 N at 5 Hz during 2.5 min. The upward bending results in a compressive strain on the top surface of the sample, which affects the crack surroundings, as can be seen in Scheme 1C. *Live/Dead staining*

Cell attachment and viability after bending were evaluated qualitatively using Live/Dead staining probes (Live Technologies). The viability was evaluated immediately after bending. Briefly, cells were incubated for 15 minutes at room temperature with a mixture of 2 μM calcein acetoxymethyl ester (Calcein AM) and 4 μM ethidium homodimer-1 (EthD-1). Images of the viable cells (green fluorescence) and dead cells (red fluorescence) were obtained using a Zeiss Axio Observer Z1 Optical Microscope with fluorescence optics.

*Cell Proliferation Assay*

Cell proliferation was quantified by measuring the activity of the living cells via mitochondrial dehydrogenase activity, by EZ4U Cell Proliferation Assay (Biomedica) according to the manufacturer's protocol. In brief, 40 μl of dye solution was added to 400 μl of medium. Then, samples were incubated for 5 hours at 37°C in a 5% $CO_2$ incubator until the sample yielded a significant increase in colour intensity. After incubation, absorbance was measured by a microplate-reader at 450 nm and subtracting background absorbance at 620 nm.

*Caspase-3/7 staining*

A CellEvent Caspase-3/7 green detection reagent was used to detect the progression of apoptosis in 15 days differentiated MC3T3-E1 cells after bending. Briefly, 8 µM of CellEvent Caspase-3/7 green detection reagent (Invitrogen) and 20 mM Hoechst 33342 nucleic acid stain (Life Technologies) were added directly to the cells in a complete osteogenic medium. Images were performed with a Zeiss Axio Observer Z1 Optical Microscope with fluorescence optics.

*Xylenol Orange staining*

Xylenol Orange powder (Sigma-Aldrich) was used to study mineralization after two weeks of differentiation in stimulated and not stimulated samples. Xyolenol Orange powder was dissolved in distillate water to make a 20mM stock solution, which was added to the ostegenic media overnight at 20 µM for detecting calcium formation. Mineralization was visualized in red using a Zeiss Axio Observer Z1 Optical Microscope with fluorescence optics. At the same time, calcein green staining was used to identify the viability and the amount of cells.

*qRT-PCR analyses*

The relative expression of the osteogenic marker gene osteocalcin (OC) at day 15 of differentiation in stimulated samples and in control samples was studied by quantitative real-time reverse transcription polymerase chain reaction (qRT-PCR). In short, the total messenger RNA (mRNA) was isolated from the samples using Maxwell RSC simply RNA extraction Kit (Promega). The isolated mRNA was reverse transcribed to cDNA with ISCRIPT Reverse Transcriptase Kit (Bio-Rad). The qRT-PCR mixture contained 50 ng cDNA, 300 nM of forward and reverse primers (Fw: "CCGGGAGCAGTGTGAGCTTA"; Rv: "AGGCGGTCTTCAAGCCATACT") and iTaq SYBR Green Supermix (Bio-Rad). The data were normalized to the expression of the housekeeping gene GAPDH. Finally, amplifications were performed in a CFX96 Real-Time PCR Detection System (Bio-Rad).

*Scanning electron microscopy*

Scanning electron microscopy (SEM) analysis was performed in cracked HA samples. Samples were previously mounted in aluminium stubs and metalized with 5nm of platinum. Images were performed with a Quanta 650 FEG microscope.

*Statistical analysis*

Statistical analyses were performed using the GraphPad Prism v6 software. Statistical significance was assessed by the two-tailed Student's t test. P-value <0.05 was considered as statistically significant. GraphPad Prism was also used to graph all the quantitative data presented as mean and standard deviations (SD). See figure legends for specific information regarding the number of biological replicates (N).

**Appendix: Analytical calculations**

2D maps of flexoelectric field for HA and $TiO_2$ presented in this paper were made from a code developed in Wolfram Mathematica. Starting from the equation of strain $\varepsilon_{ij}^{el}$ around a crack mode I [34]:

$$\varepsilon_{ij}^{el} = \frac{1+\upsilon}{Y}\sigma_{ij} - 3\frac{\upsilon}{Y}\sigma_m \delta_{ij}, \qquad (1)$$

where $\sigma_{ij}$ is the stress applied to the crack in each direction, $\upsilon$ is the Poisson ratio, Y the Young's modulus, $\sigma_m = \frac{\sigma_{11}+\sigma_{22}+\sigma_{12}}{3}$ is the mean stress and $\delta_{ij}$ is Kronecker delta function. Furthemore, for a crack mode I, the stresses in each direction are given by:

$$\sigma_{11} = \frac{K_I}{Y_{11}\sqrt{2\pi r}} \cos\frac{\theta}{2}\left(1 - \sin\frac{\theta}{2}\sin\frac{3\theta}{2}\right) \qquad (2)$$

$$\sigma_{22} = \frac{K_I}{Y_{22}\sqrt{2\pi r}} \cos\frac{\theta}{2}\left(1 + \sin\frac{\theta}{2}\sin\frac{3\theta}{2}\right) \qquad (3)$$

$$\tau_{12} = \frac{K_I}{Y_{12}\sqrt{2\pi r}} \cos\frac{\theta}{2}\sin\frac{\theta}{2}\cos\frac{3\theta}{2}, \qquad (4)$$

where $K_I$ is the intensity factor of the material and $Y_{ij}$ is the Young's modulus in the different directions, in these cases both materials were considered isotropic. These stresses are the maximum that the crack can withstand without propagating.

The stress equations were transformed into Cartesian coordinates in order to compute the strain (this is necessary to compute the flexoelectric field), and then the partial derivates of the strain in each direction were taken to calculate the strain gradient. The different components of the flexoelectric field were obtained from the following equations [19]:

$$E_1 = f_{11} \frac{\partial \varepsilon_{11}}{\partial x_1} + f_{12} \frac{\partial \varepsilon_{22}}{\partial x_1} \tag{5}$$

$$E_2 = f_{22} \frac{\partial \varepsilon_{22}}{\partial x_2} + f_{21} \frac{\partial \varepsilon_{11}}{\partial x_2} \tag{6}$$

$$E = \sqrt{E_1^2 + E_2^2}, \tag{7}$$

where, $f_{ij}$ is the flexocoupling tensor, the factors $\frac{\partial \varepsilon_{ij}}{\partial x_k}$ are the strain gradients and E is the magnitude of the electric field and the value ploted in the maps ($E_1$, $E_2$ are the components of the electric field along the x and y directions respectively). The flexocoupling tensor was calculated as the product of the effective flexoelectric coefficient $\mu_{eff}$ and the dielectric constant $\epsilon$, both values measured with the same methodology previously reported for bone [18]:

$$\mu_{eff} = f_{eff}\epsilon. \tag{8}$$

For this calculation, $f_{11} = f_{22} = f_{12} = f_{21} = f_{eff}$, and the shear component was taken as null. This approximation is common because the shear coefficient is poorly characterized [18, 35]. The results are thus correct within the order-of-magnitude approximation, which is the common case for flexoelectricity as it is impossible to measure independently all the components of the flexoelectric tensor [19].

In the case of $TiO_2$ the critical intensity factor used was 2.4 M Pa m$^{1/2}$ and the Poisson ratio was 0.27 [36], the Young's modulus used was 230 GPa and the flexocoupling value was 4 V, both from measurements made in our laboratory. Meanwhile, for hydroxyapatite, the critical intensity factor was 1 MPa m$^{1/2}$, the Poisson ratio used was 0.2 [37], the Young's modulus was 80 GPa and the flexocoupling coefficient 11 V, the last two values obtained from our measurements.

**References**

1. Hazenberg, J. G. *et al.* Microdamage detection and repair in bone: fracture mechanics, histology, cell biology. *Technology and health care : official journal of the European Society for Engineering and Medicine* **17**, 67-75, doi:10.3233/THC-2009-0536 (2009).

2. Mori, S. & Burr, D. B. Increased intracortical remodeling following fatigue damage. *Bone* **14**, 103-109 (1993).

3. O'Brien, F. J., Taylor, D. & Lee, T. C. Microcrack accumulation at different intervals during fatigue testing of compact bone. *Journal of biomechanics* **36**, 973-980 (2003).

4. Estberg, L. *et al.* High-speed exercise history and catastrophic racing fracture in thoroughbreds. *Am J Vet Res* **57**, 1549-1555 (1996).

5. Muir, P., Johnson, K. A. & Ruaux-Mason, C. P. In vivo matrix microdamage in a naturally occurring canine fatigue fracture. *Bone* **25**, 571-576, doi:10.1016/s8756-3282(99)00205-7 (1999).

6. Burr, D. B. Bone, exercise, and stress fractures. *Exerc Sport Sci Rev* **25**, 171-194 (1997).

7. Zarrinkalam, K. H., Kuliwaba, J. S., Martin, R. B., Wallwork, M. A. & Fazzalari, N. L. New insights into the propagation of fatigue damage in cortical bone using confocal microscopy and chelating fluorochromes. *Eur J Morphol* **42**, 81-90 (2005).

8. Wegst, U. G., Bai, H., Saiz, E., Tomsia, A. P. & Ritchie, R. O. Bioinspired structural materials. *Nat Mater* **14**, 23-36, doi:10.1038/nmat4089 (2015).

9. Verborgt, O., Gibson, G. J. & Schaffler, M. B. Loss of osteocyte integrity in association with microdamage and bone remodeling after fatigue in vivo. *Journal of bone and mineral research : the official journal of the American Society for Bone and Mineral Research* **15**, 60-67, doi:10.1359/jbmr.2000.15.1.60 (2000).

10. Komori, T. Functions of the osteocyte network in the regulation of bone mass. *Cell Tissue Res* **352**, 191-198, doi:10.1007/s00441-012-1546-x (2013).

11. Sun, X. *et al.* Mechanical stretch induced calcium efflux from bone matrix stimulates osteoblasts. *Bone* **50**, 581-591, doi:10.1016/j.bone.2011.12.015 (2012).

12. Shu, Y. *et al.* Surface microcracks signal osteoblasts to regulate alignment and bone formation. *Materials science & engineering. C, Materials for biological applications* **44**, 191-200, doi:10.1016/j.msec.2014.08.036 (2014).

13. Aquino-Martinez, R., Artigas, N., Gamez, B., Rosa, J. L. & Ventura, F. Extracellular calcium promotes bone formation from bone marrow mesenchymal stem cells by amplifying the effects of BMP-2 on SMAD signalling. *PloS one* **12**, e0178158, doi:10.1371/journal.pone.0178158 (2017).

14. Blair, H. C. *et al.* Calcium and bone disease. *Biofactors* **37**, 159-167, doi:10.1002/biof.143 (2011).

15. Bassett, C. A. & Becker, R. O. Generation of electric potentials by bone in response to mechanical stress. *Science* **137**, 1063-1064 (1962).

16. Fukada, E. & Yasuda, I. Piezoelectric Effects in Collagen. *Jpn. J. Appl. Phys.* **3**, 502B–502B (1964).

17. Fukada, E. & Yasuda, I. On the Piezoelectric Effect of Bone. *Journal of the Physical Society of Japan* **12**, 1158-1162, doi:10.1143/JPSJ.12.1158 (1957).

18. Vasquez-Sancho, F., Abdollahi, A., Damjanovic, D. & Catalan, G. Flexoelectricity in Bones. *Advanced materials* **30**, doi:10.1002/adma.201705316 (2018).

19. Zubko, P., Catalan, G. & Tagantsev, A. K. Flexoelectric Effect in Solids. *Annual Review of Materials Research* **43**, 387-421, doi:10.1146/annurev-matsci-071312-121634 (2013).

20. Cross, L. E. Flexoelectric effects: Charge separation in insulating solids subjected to elastic strain gradients. *Journal of Materials Science* **41**, 53-63, doi:10.1007/s10853-005-5916-6 (2006).

21. McCullen, S. D. *et al.* Application of low-frequency alternating current electric fields via interdigitated electrodes: effects on cellular viability, cytoplasmic calcium, and osteogenic differentiation of human adipose-derived stem cells. *Tissue engineering. Part C, Methods* **16**, 1377-1386, doi:10.1089/ten.TEC.2009.0751 (2010).

22. Cordero-Edwards, K., Kianirad, H., Canalias, C., Sort, J. & Catalan, G. Flexoelectric Fracture-Ratchet Effect in Ferroelectrics. *Phys Rev Lett* **122**, 135502, doi:10.1103/PhysRevLett.122.135502 (2019).

23. Noble, B. Bone microdamage and cell apoptosis. *European cells & materials* **6**, 46-55; discusssion 55 (2003).

24. Jilka, R. L., Noble, B. & Weinstein, R. S. Osteocyte apoptosis. *Bone* **54**, 264-271, doi:10.1016/j.bone.2012.11.038 (2013).

25. Bentolila, V. *et al.* Intracortical remodeling in adult rat long bones after fatigue loading. *Bone* **23**, 275-281 (1998).

26. Mera, P. *et al.* Osteocalcin Signaling in Myofibers Is Necessary and Sufficient for Optimum Adaptation to Exercise. *Cell metabolism* **23**, 1078-1092, doi:10.1016/j.cmet.2016.05.004 (2016).

27. Aaron, R. K., Ciombor, D. M., Wang, S. & Simon, B. Clinical biophysics: the promotion of skeletal repair by physical forces. *Annals of the New York Academy of Sciences* **1068**, 513-531, doi:10.1196/annals.1346.045 (2006).

28. Lavine, L. S. & Grodzinsky, A. J. Electrical stimulation of repair of bone. *The Journal of bone and joint surgery. American volume* **69**, 626-630 (1987).

29. Sun, S., Liu, Y., Lipsky, S. & Cho, M. Physical manipulation of calcium oscillations facilitates osteodifferentiation of human mesenchymal stem cells. *FASEB journal : official publication of the Federation of American Societies for Experimental Biology* **21**, 1472-1480, doi:10.1096/fj.06-7153com (2007).

30. Sun, S., Titushkin, I. & Cho, M. Regulation of mesenchymal stem cell adhesion and orientation in 3D collagen scaffold by electrical stimulus. *Bioelectrochemistry* **69**, 133-141, doi:10.1016/j.bioelechem.2005.11.007 (2006).

31. Titushkin, I. A., Rao, V. S. & Cho, M. R. Mode- and cell-type dependent calcium responses induced by electrical stimulus. *IEEE Transactions on Plasma Science* **32**, 1614-1619 (2004).

32. Piacentini, R., Ripoli, C., Mezzogori, D., Azzena, G. B. & Grassi, C. Extremely low-frequency electromagnetic fields promote in vitro neurogenesis via upregulation of Ca(v)1-channel activity. *Journal of cellular physiology* **215**, 129-139, doi:10.1002/jcp.21293 (2008).

33. Ooi, C. Y., Hamdi, M. & Ramesh, S. Properties of hydroxyapatite produced by annealing of bovine bone. *Ceramics International* **33**, 1171-1177, doi:10.1016/j.ceramint.2006.04.001 (2007).

34. Lawn, B. Fracture of brittle solids. *Cambridge University Press* (1993).

35. Abdollahi, A., Peco C., Millán D., Arroyo, M. & Arias, I. Computational evaluation of the flexoelectric effect in dielectric solids. *Journal of Applied Physics* **116**, 093502 (2014).

36. AZoM. Titanium Dioxide - Titania (TiO2), Titanium Dioxide -Titania (TiO2) (2002).

37. Currey, J. D. Bones: Structure and Mechanics. *Princeton University Press* (2002).

**Acknowledgments**

This research was funded by an ERC starting grant (grant number ERC 308023). We acknowledge additional support of the Generalitat de Catalunya (Grant 2017 SGR 579) and the European Union's Horizon 2020 Programme, under the Marie Curie Marie Sklodowska-Curie grant agreement No 753186, "Flexobonegraft". All research at ICN2 receives the support of the institute's Severo Ochoa Excellence Award (MINECO, grant no. SEV-2017-0706). F. Vasquez-Sancho acknowledge CONICIT, MICIT and University of Costa Rica for their support during the doctorate program. We acknowledge Rafael León (Mechanical workshop, ICN2) for the design and fabrication of the crack generation and sample stimulation systems.

**Author contributions**

R. N-T and F. V-S performed most of the experimental work and contributed to experimental design, data analysis, discussions, and writing the paper. N. B helped with the experiments. G. C conceived and supervised the study and co-wrote the paper.

**Competing interests**

The authors declare no competing interests.